\documentstyle[epsfig,12pt,english,here,oldlfont]{article}
\begin{document}
\baselineskip=0.8cm
\newcommand{\ini}{\begin{equation}}
\newcommand{\fin}{\end{equation}}
\newcommand{\inir}{\begin{eqnarray}}
\newcommand{\finr}{\end{eqnarray}}
\newcommand{\inif}{\begin{figure}}
\newcommand{\finf}{\end{figure}}
\newcommand{\bc}{\begin{center}}
\newcommand{\ec}{\end{center}}
\def\ol{\overline}
\def\pa{\partial}
\def\ra{\rightarrow}
\def\ts{\times}
\def\df{\dotfill}
\def\bs{\backslash}

$~$

\hfill DSF-T-98/41

\vspace{1 cm}

\centerline{\LARGE{Lepton mass matrices from}}

\centerline{\LARGE{a quark-lepton analogy}}

\vspace{1 cm}

\centerline{\large{D. Falcone}}

\vspace{1 cm}

\centerline{Dipartimento di Scienze Fisiche, Universit\`a di Napoli,}
\centerline{Mostra d'Oltremare, Pad. 19, I-80125, Napoli, Italy,}
\centerline{and INFN, Sezione di Napoli, Napoli, Italy.}

\centerline{e-mail: falcone@na.infn.it}

\vspace{1 cm}

\centerline{ABSTRACT}

We study the implications of having a similarity between quark and lepton
mixing in the Dirac sector of the Standard Model plus the right-handed
neutrino. This enable us to describe all masses and mixings in the Dirac sector
in terms of only five parameters: three mass scales $m_b$, $m_{\tau}$ and
$m_t$, one parameter $\lambda$ describing all the mixings, and a CP violating
phase in the quark sector. Then, from experimental data on neutrino masses and
mixings, we extract the heavy neutrino mass matrix. The approach considered,
although does not contraddict Grand Unified Theories, can help to find other
theoretical models of fermion masses.

PACS numbers: 12.15.Ff, 14.60.Pq

\newpage

The Super-Kamiokande Collaboration has confirmed the oscillation of atmospheric
neutrinos \cite{skc}. This evidence, as well as the indications of oscillation of
solar neutrinos to solve the solar neutrino problem \cite{dgp},
leads to a finite mass of neutrinos.
However, in the Minimal Standard Model (MSM), neutrino is massless.
In fact, with only the left-handed neutrino
$\nu_L$ we cannot get a Dirac mass, and with only the Higgs
doublet we cannot get a Majorana mass for $\nu_L$.
Adding a right-handed neutrino $\nu_R$ allows to build a Dirac mass term of
neutrino, in analogy with the other Dirac masses of fermions in the theory.
But neutrino mass is very small
if compared with the other fermion masses. The see-saw mechanism \cite{ssm}
explains this feature, giving a large Majorana mass to $\nu_R$.
This mass is not constrained and in fact defines a new scale.

Regarding the mixings, recent data imply large mixing between second and third
lepton family \cite{skc}, while the mixing between first and second lepton
family may be large or small \cite{hl}. On the contrary, in the quark sector
all mixings are small \cite{pdg}. The see-saw mechanism helps to understand
also these features: the Dirac sector of the theory may give similar
quark and lepton mixings, while the effect of the Majorana sector can
enhance the effective lepton mixing \cite{yus}.

Motivated by that, we construct explicitly a quark-lepton analogy in the Dirac
sector, suggested by ref.\cite{ft}, and derive all fermion mass matrices,
using experimental limits on neutrino masses and mixings \cite{bg}. A very
simple scheme comes out. This scheme has some resemblance with the Grand
Unified Theory (GUT) scheme \cite{bky} (see also \cite{rrr}),
but we work it out within a more general
approach, where the MSM is enlarged to include
the right-handed neutrino only.

Now, we first summarize the experimental data on lepton mixing, then we briefly
describe the formalism of the see-saw mechanism, and finally explore the
consequences of a quark-lepton analogy in the Dirac mixing of the theory.

Weak and mass eigenstates of neutrino are connected by the relation
\ini
\nu_{\alpha}=U_{\alpha i} \nu_i
\fin
where $U$ is a unitary matrix, $\alpha=e,\mu,\tau$, and $i=1,2,3$.
According to ref.\cite{bg} we write $U$ as
\ini
U=\left(\begin{array}{ccc}
       c_{12} & s_{12} & 0 \\
     -s_{12}c_{23} & c_{12}c_{23} & s_{23} \\
      s_{12}s_{23}  & -c_{12}c_{23} & c_{23}
    \end{array}\right).
\fin
(We put zero in position 1-3, although it is only constrained to be much less
than one.) 
The experimental data on oscillation of atmospheric and solar neutrinos lead
to three possible numerical forms for $U$.
These correspond to the three solutions of
the solar neutrino problem, namely small mixing or large mixing MSW
\cite{msw}, and
vacuum oscillations. The small mixing solution is now preferred over the
large mixing one \cite{flm}, and we refer to this solution and the vacuum
oscillations, although the same calculation can be done for the large mixing
MSW. Choosing the central values of $s_{12}$ and $s_{23}$
of ref.\cite{bg}, we obtain, for small mixing MSW,
\ini
U = \left(\begin{array}{ccc}
       1 & 0.04 & 0 \\
     -0.032 & 0.80 & 0.60 \\
      0.024 & -0.60 & 0.80
    \end{array}\right)
\fin
and, for vacuum oscillations,
\ini
U = \left(\begin{array}{ccc}
       0.80 & 0.60 & 0 \\
     -0.474 & 0.632 & 0.61 \\
      0.366 & -0.488 & 0.79
    \end{array}\right).
\fin
In the following we need also the values of light neutrino masses.
From the experimental limits on $\Delta m_{32}^2$, $\Delta m_{21}^2$ we take,
for example,
\ini
m_{\nu_1}=2.2 \times 10^{-4},~m_{\nu_2}=2.8 \times 10^{-3},
~m_{\nu_3}=3.6 \times 10^{-2}~~(eV)
\fin
and
\ini
m_{\nu_1}= 3.0 \times 10^{-9},~m_{\nu_2}=1.0 \times 10^{-5},
~m_{\nu_3}=3.6 \times 10^{-2}~~(eV),
\fin
for small mixing MSW and vacuum oscillations,
respectively. We assume a hierarchical pattern
$m_{\nu_1} \ll m_{\nu_2} \ll m_{\nu_3}$, which gives $m_{\nu_3}$, $m_{\nu_2}$,
and $m_{\nu_1}/m_{\nu_2}=m_{\nu_2}/m_{\nu_3}$, which gives $m_{\nu_1}$. 

Now we briefly explain the formalism of the Standard Model plus $\nu_R$ with the
see-saw mechanism. The part of the Lagrangian we are interested in is
\ini
\ol{e}_L M_e e_R+\ol{\nu}_L M_{\nu} \nu_R+g \ol{\nu}_L e_L W+\ol{\nu}^c_L M'_R
\nu_R
\fin
where $M_e$ and $M_{\nu}$ are the Dirac mass matrices of charged leptons and
neutrinos respectively, and $M'_R$ is the Majorana mass matrix of right-handed
neutrinos. We assume the elements of $M'_R$ much greater than those of
$M_{\nu}$.
When we diagonalize the Dirac 
mass matrices we have (renaming the fermion fields)
\ini
\ol{e}_L D_e e_R+\ol{\nu}_L D_{\nu} \nu_R+g \ol{\nu}_L V_D e_L W
+\ol{\nu}^c_L M_R \nu_R
\fin
where $D_e$ and $D_{\nu}$ are diagonal and $V_D$ is the analogous of $V_{CKM}$
\cite{ckm} in the sense that it rises from the diagonalization of the
Dirac lepton sector. The see-saw mechanism leads to the effective Lagrangian  
\ini
\ol{e}_L D_e e_R+\ol{\nu}_L M_L {\nu}^c_R+g \ol{\nu}_L V_D e_L W
+\ol{\nu}^c_L M_R \nu_R
\fin
with the Majorana mass matrix of left-handed neutrinos
\ini
M_L=D_{\nu} M_R^{-1} D_{\nu}.
\fin  
Now, if we diagonalize also $M_L$, we obtain
\ini
\ol{e}_L D_e e_R+\ol{\nu}_L D_L {\nu}^c_R+g \ol{\nu}_L V_s V_D e_L W+
\ol{\nu}^c_L M_R \nu_R
\fin
where the unitary matrix $V_s$ specifies the effect of the see-saw mechanism
on lepton mixing that is the effect of the Majorana mass matrix of the
right-handed neutrino \cite{yus1}. The product
\ini
V_{lep}=V_s V_D
\fin
is the lepton mixing matrix that appears in the charged current interaction, and
is related to the neutrino mixing matrix $U$ by
\ini
V_{lep}=U^+.
\fin     
As for the quark sector, we have now
\ini
V_D=V_{\nu}^+ V_e,
\fin
which is the analogous of $V_{CKM}=V_u^+ V_d$, while
\ini
V_s M_L V_s^+=D_L
\fin
where $D_L=diag(m_{\nu_1},m_{\nu_2},m_{\nu_3})$ gives the masses of light
neutrinos.

In the quark sector we can always choose $M_u$ diagonal and $M_d$ with three
zeros in certain positions \cite{ft,fprhs}.
In the same way, in the Dirac lepton sector,
we can always choose $M_{\nu}$ diagonal and $M_e$
with three zeros in the same positions  (then $M'_R=M_R$).
For the quark case, using one of these bases,
in ref.\cite{ft}, a very simple description of the down quark mass matrix
$M_d$ and the $V_{CKM}$ matrix was inferred:
\ini
M_d \simeq \left(\begin{array}{ccc}
       0 & \sqrt{m_d m_s} & 0 \\
     \sqrt{m_d m_s} & m_s & m_s \\
      0 & m_b/\sqrt{5} & 2m_b/\sqrt{5}
    \end{array}\right)
\fin
that yields
\ini
V_{us} \simeq \sqrt{\frac{m_d}{m_s}},~V_{cb} \simeq \frac{3}{\sqrt{5}}
\frac{m_s}{m_b},~
V_{ub} \simeq \frac{1}{\sqrt{5}}\frac{\sqrt{m_d m_s}}{m_b}.
\fin
Such a basis shows, in a
single matrix, $M_d$, the hierarchy of both down quark masses and flavor
mixings. As also charged lepton masses are hierarchical, one can imagine a
similar structure of mixing in the Dirac lepton sector and in the quark sector. 
Then, we set
\ini
V_D \simeq \left(\begin{array}{ccc}
       1-\frac{m_e}{m_{\mu}} & \sqrt{\frac{m_e}{m_{\mu}}} & \frac{1}{\sqrt{5}} 
       \frac{\sqrt{m_e m_{\mu}}}{m_\tau} \\
     -\sqrt{\frac{m_e}{m_{\mu}}} & 1-\frac{m_e}{m_{\mu}} & \frac{3}{\sqrt{5}} 
      \frac{m_{\mu}}{m_{\tau}} \\
      \frac{2}{\sqrt{5}}\frac{\sqrt{m_e m_{\mu}}}{m_\tau} & -\frac{3}{\sqrt{5}}
     \frac{m_{\mu}}{m_{\tau}} & 1
    \end{array}\right)
\fin
and hence, in analogy with ref.\cite{ft}, we can write
\ini
M_e \simeq \left(\begin{array}{ccc}
       0 & \sqrt{m_e m_{\mu}} & 0 \\
     \sqrt{m_e m_{\mu}} & m_{\mu} & m_{\mu} \\
      0 & m_{\tau}/\sqrt{5} & 2m_{\tau}/\sqrt{5}
    \end{array}\right)
\fin
(we set the CP violating phase in the leptonic sector equal to zero),
while $M_{\nu}$ is diagonal.
Using quark and lepton masses (at the
scale $M_Z$) as calculated in ref.\cite{fk}, one can check from Eqns.(16),(19)
that
\ini
M_d \simeq m_b \left(\begin{array}{ccc}
       0 & \lambda^3 & 0 \\
     \lambda^3 & \lambda^2 & \lambda^2 \\
      0 & \frac{1}{\sqrt{5}} & \frac{2}{\sqrt{5}}
    \end{array}\right),
\fin
\ini
M_e \simeq m_{\tau} \left(\begin{array}{ccc}
       0 & \frac{1}{2}\lambda^3 & 0 \\
     \frac{1}{2}\lambda^3 & \frac{3}{2}\lambda^2 & \frac{3}{2}\lambda^2 \\
      0 & \frac{1}{\sqrt{5}} & \frac{2}{\sqrt{5}}
    \end{array}\right),
\fin
with $\lambda \simeq \sqrt{m_d/m_s} \simeq \sqrt{m_s/m_b}$.
This means that, on the basis of
ref.\cite{ft} (both $M_u$ and $M_{\nu}$ are diagonal), if we assume Eqn.(18),
then we
obtain a simple description of both $M_d$ and $M_e$ in terms of three real
parameters plus one phase in $M_d$ (not reported). Actually we have
\ini
V_{D12} \simeq \frac{1}{3}\lambda,~
\fin
\ini  
V_{D23} \simeq \frac{9}{2\sqrt{5}}\lambda^2 \simeq 2\lambda^2,~  
\fin
\ini
V_{D13} \simeq \frac{1}{2\sqrt{5}}\lambda^3 \simeq \lambda^4.
\fin  
Now we proceed towards the determination of Majorana mass matrices.
From Eqn.(12) we have
\ini
V_s=V_{lep} V_D^+
\fin
from which, by Eqn.(13), we can calculate $V_s$. Then, from Eqn.(15) we get
\ini
M_L=V_s^+ D_L V_s,
\fin 
and from Eqn.(10) we have also
\ini
M_R=D_{\nu} M_L^{-1} D_{\nu}.
\fin
As $M_d$ is almost proportional to $M_e$ by the factor $m_b/m_{\tau}$, we can
assume
\ini
D_{\nu} \simeq \frac{m_{\tau}}{m_b} D_u,
\fin
which gives the Dirac masses of neutrinos $m_1=0.00136, m_2=0.394, m_3=105$
(GeV). In GUT's the ratio $m_b/m_{\tau}$ is related to renormalization
group evolution of Yukawa couplings from the intermediate scale,
rather than the unification scale \cite{sspu}, to the weak scale.
We have also $\sqrt{m_u/m_t}\simeq m_c/m_t \simeq \lambda^4$ \cite{hw}, thus
\ini
M_u=D_u \simeq m_t~diag(\lambda^8,\lambda^4,1),
\fin
and then all  Dirac masses and mixings are expressed in terms of four real
parameters, that is $m_b$, $m_{\tau}$, $m_t$, and $\lambda$.
By Eqn.(27) we can now calculate the heavy neutrino mass
matrix $M_R$. We yield (in GeV) for the two cases
of small mixing MSW and vacuum oscillations, respectively:
\ini
M_R = \left(\begin{array}{ccc} 
     8.3 \times 10^{6} & -2.2 \times 10^{8} & 1.6 \times 10^{10} \\
     -2.2 \times 10^{8} & 3.9 \times 10^{10} & -7.2 \times 10^{12} \\
     1.6 \times 10^{10} & -7.2 \times 10^{12} & 1.9 \times 10^{15}
    \end{array}\right),
\fin
with eigenvalues
\ini
M_1 = 6.0 \times 10^6,M_2 = 1.2 \times 10^{10},M_3 = 1.9 \times 10^{15},
\fin
and
\ini
M_R = \left(\begin{array}{ccc}
     3.6 \times 10^{11} & -6.8 \times 10^{13} & 1.5 \times 10^{16} \\
     -6.8 \times 10^{13} & 1.3 \times 10^{16} & -2.8 \times 10^{18} \\
     1.5 \times 10^{16} & -2.8 \times 10^{18} & 6.0 \times 10^{20}
    \end{array} \right),
\fin
with eigenvalues
\ini
M_1 = 2.6 \times 10^7,M_2 = 1.9 \times 10^{11},M_3 = 6.0 \times 10^{20}.
\fin
The highest mass scale appearing is near the grand unification scale, in the
former case, while it is near the super unification scale
(the Planck mass), in the latter. We also observe a huge hierarchy.
However, $M_R$ depends on values of light neutrino masses, which are not well
estabilished. Therefore, if the origin of the right-handed neutrino mass is at
some intermediate scale, then, from Eqns.(30),(32), we find that the small
mixing MSW solution is favoured.

Now we compare our scheme with the one in ref.\cite{bky}, based on a GUT (for
example $SO(10)$). There, mass matrices are simmetric:
\ini
M_d \simeq m_b \left(\begin{array}{ccc}
       0 & \lambda^3 & 0 \\
     \lambda^3 & \lambda^2 & \lambda^2 \\
      0 & \lambda^2 & 1
    \end{array}\right),
\fin
\ini
M_e \simeq m_{\tau} \left(\begin{array}{ccc}
       0 & \lambda^3 & 0 \\
     \lambda^3 & -3 \lambda^2 & \lambda^2 \\
      0 & \lambda^2 & 1
    \end{array}\right),
\fin
\ini
M_u \simeq m_t \left(\begin{array}{ccc}
       0 & 0 & \lambda^4 \\
       0 & \lambda^4 & 0 \\
       \lambda^4 & 0 & 1
    \end{array}\right),
\fin
\ini
M_{\nu} \simeq \frac{m_{\tau}}{m_b} M_u,
\fin
and lead to
\ini
V_{D12}\simeq \frac{1}{3}\lambda,
\fin
\ini
V_{D23}\simeq \lambda^2,
\fin
\ini
V_{D13}\simeq \lambda^4.
\fin
The dependence of mixings on $\lambda$ is similar to our scheme, but
Eqns.(32-35) rely on some $ad~hoc$ couplings of Higgs bosons to fermions
(for example, in $SO(10)$, the coefficient $-3$ should come from a {\bf 126}
which couples only to the second generation \cite{gj}, while the other
couplings come from a {\bf 10}). We argue that
Eqns. (20),(21), with $M_u$ and
$M_{\nu}$ diagonal, should approximately be the expression of Eqns.(34)-(37)
on our basis. Only $V_{23}$ in Eqn.(23) is twice the value in Eqn.(39).

A further remark can be done:
if we change the numerical coefficients in $V_D$ (Eqn.(18)),
leaving the same dependence on mass ratios,
the result of a few-parameter description of mass and mixing does not change.
This suggests also an intriguing possibility, that is $V_D=V_{CKM}$,
which differs from GUT mixings for $V_{D12}\simeq \lambda$,
and should imply a CP violating phase in the leptonic sector too.
A better experimental knowledge of fermion masses and mixings,
as well as theoretical
speculations \cite{ilr}, can decide among different patterns of lepton mass and
mixing in the Dirac sector. Using our basis should help this task.                                 

In conclusion, from experimental data on neutrino mass and
mixing, and a quark-lepton analogy, we obtained all lepton mass matrices.
The ideas developed here, although could
be in agreement with GUT's, can also be useful to study some
other models of fermion masses and mixings.

We thank F. Buccella for helpful comments, and L. Rosa, O. Pisanti,

F. Tramontano, S. Esposito for discussions.

\end{document}